\documentstyle[aas2pp4]{article}
\newcommand{\cgro}{{\it CGRO}}
\newcommand{\vvmax}{\mbox{$\langle V/V_{max} \rangle$}}

\begin{document}
\title{The Intensity Distribution of Faint Gamma-Ray Bursts Detected
  with BATSE}

\author{Jefferson M. Kommers,\altaffilmark{1} Walter H. G.
  Lewin,\altaffilmark{1} Chryssa Kouveliotou,\altaffilmark{2,3} Jan van
  Paradijs,\altaffilmark{4,5} Geoffrey N. Pendleton,\altaffilmark{4}
  Charles A. Meegan,\altaffilmark{3} and Gerald J. Fishman\altaffilmark{3}}

\altaffiltext{1}{Department of Physics and Center for Space Research,
  Massachusetts Institute of Technology, Cambridge, MA 02139; {\tt kommers@space.mit.edu}}
\altaffiltext{2}{Universities Space Research Association, Huntsville, AL 35800}
\altaffiltext{3}{NASA/Marshall Space Flight Center, Huntsville, AL
  35812}
\altaffiltext{4}{University of Alabama in Huntsville, Huntsville, AL
  35812}
\altaffiltext{5}{University of Amsterdam, Amsterdam, Netherlands}

\slugcomment{To appear in ApJ}

\begin{abstract}
  We have recently completed a search of 6 years of archival BATSE
  data for gamma-ray bursts (GRBs) that were too faint to activate the
  real-time burst detection system running onboard the spacecraft.
  These ``non-triggered'' bursts can be combined with the
  ``triggered'' bursts detected onboard to produce a GRB intensity
  distribution that reaches peak fluxes a factor of $\sim 2$ lower
  than could be studied previously.  The value of the $\langle
  V/V_{max}\rangle$ statistic (in Euclidean space) for the bursts we
  detect is $0.177 \pm 0.006$.  This surprisingly low value is
  obtained because we detected very few bursts on the 4.096 s and
  8.192 s time scales (where most bursts have their highest
  signal-to-noise ratio) that were not already detected on the 1.024 s
  time scale.  If allowance is made for a power-law distribution of
  intrinsic peak luminosities, the extended peak flux distribution is
  consistent with models in which the redshift distribution of the
  gamma-ray burst rate approximately traces the star formation history
  of the Universe.  We argue that this class of models is preferred
  over those in which the burst rate is independent of redshift.  We
  use the peak flux distribution to derive a limit of 10\% (99\%
  confidence) on the fraction of the total burst rate that could be
  contributed by a spatially homogeneous (in Euclidean space)
  subpopulation of burst sources, such as type Ib/c supernovae.  These
  results lend support to the conclusions of previous studies
  predicting that relatively few faint ``classical'' GRBs will be
  found below the BATSE onboard detection threshold.
\end{abstract}


\section{Introduction}
\setcounter{footnote}{0}
\label{sec:intro}
The origin of some, and possibly all, gamma-ray bursts (GRBs) at
cosmological distances has been firmly established with the
identification of X-ray, optical, and radio afterglows
(\cite{Costa97}; \cite{jvp97}; \cite{Frail97}) and the subsequent
measurement of cosmological redshifts for at least four of the optical
afterglows and/or their host galaxies (\cite{Metzger97};
\cite{Kulkarni98a}; \cite{Djorgovski99}; \cite{Djorgovski98}).  The
objects responsible for producing the majority of GRBs, the gamma-ray
bursters themselves, have yet to be understood, however.  To obtain an
understanding of the spatial distribution of sources and the
distribution of their burst luminosities is a crucial step towards
identifying the physical processes that produce GRBs.

Before the rapid follow-up of GRB afterglows was made possible by the
{\it BeppoSAX\/} satellite, the only way to test hypotheses about the
spatial and luminosity distributions was to fit parametric models to
the measured characteristics of the bursts themselves.  For this
purpose the distribution of GRB intensities was used (see, for
example, \cite{Fenimore93}; \cite{Rutledge95}; \cite{Fenimore95};
\cite{Cohen95}; \cite{Hakkila96}; and references therein).  The
effects of cosmological time dilation on the time profiles of bright
versus faint bursts were also studied (\cite{Norris95}).  Since
optical spectroscopic redshifts are so far associated with only four
(possibly five) bursts\footnote{The proposed association of GRB 980425
  with SN 1998bw (z = 0.008; Galama et al.\ 1998) may indicate a
  separate class of GRBs (Bloom et al.\ 1998).  We will therefore
  consider that event separately (see section 3.2).}, number counts as
a function of intensity remain an important tool for exploring the
possible spatial and luminosity distributions of GRBs.

Several recent papers (Totani 1997, 1998; \cite{Wijers98};
\cite{Krum98}; \cite{Mao98}) have used the observed GRB intensity
distributions to investigate the possibility that the redshift
distribution of gamma-ray bursters traces the global star formation
history of the Universe.  The motivation for this hypothesis is a
collection of theoretical models in which GRBs are produced by stellar
objects that evolve from their formation to their bursting phase on a
time scale of $\sim 100$ Myr or less.  This group of models includes
the merging of a neutron star with another neutron star or a black
hole, the collapse of a massive star, and the collapse of a
Chandrasekhar-mass white dwarf (see \cite{Wijers98} for references).
In these scenarios, the cosmological redshift distribution of the GRB
rate should approximately follow the redshift distribution of the
formation rate of stellar objects; in other words, the GRB rate should
trace the global star formation history of the Universe.  This
hypothesis appears to solve some puzzling aspects of the observations,
such as the ``no host'' problem (\cite{Schaefer97}; \cite{Wijers98}).

The star formation rate (SFR) as a function of redshift has been
studied by Lilly et al.\ (1996), Fall, Charlot, \& Pei (1996), Madau,
Pozzetti, \& Dickinson (1998), and Hughes et al.  (1998).  The
principal result of these studies is that the SFR was substantially
higher in the past.  Between the present and $z \approx 1$ the SFR
increases by a factor of $\sim 10$; it peaks somewhere in the range $z
\approx 1$ to $z \approx 3$; and it decreases to a rate comparable to
the present by $z \approx$ 4--5 (this last point remains uncertain).

Totani (1997), Wijers et al.\ (1998), Krumholz et al.\ (1998), and Mao
\& Mo (1998) all find that the GRB peak flux number counts can
accommodate the hypothesis that the GRB rate follows the SFR.  Among
the important conclusions that these authors derive from this
interpretation of the data are the following: 1) that the faintest
gamma-ray bursts observed with the Burst and Transient Source
Experiment (BATSE) onboard the {\it Compton Gamma Ray Observatory\/}
(\cgro) are already produced at redshifts of $z \approx 3$ to $z
\approx 6$ (\cite{Wijers98}; but see Section~\ref{sec:discuss}); and
2) that more sensitive experiments are unlikely to discover large
numbers of faint GRBs (of the ``classical'' kind that are detected
with current instruments) below the BATSE onboard detection threshold.
The latter conclusion has important implications for the design and
operation of future GRB detectors, which will test the behavior of GRB
number counts at intensities well below the BATSE threshold.

We have recently completed a search of 6 years of archival data from
BATSE for GRBs and other transients that did {\it not\/} activate the
real-time burst detection system (or ``trigger'') running onboard the
spacecraft.  A GRB or other transient may fail to activate the BATSE
onboard burst trigger for any of several reasons.  The burst may be
too faint to exceed the onboard detection threshold, it may occur
while the onboard trigger is disabled for technical reasons, it may
occur while the onboard trigger is optimized for detecting non-GRB
phenomena, or it may artificially raise the onboard background
estimate and be mistaken for a below-threshold event.  Our search of
the archival data is sensitive to GRBs with peak fluxes (measured over
1.024 s in the 50--300 keV energy range) that are a factor of $\sim 2$
lower than can be detected with the onboard trigger in its nominal
configuration.  Thus our search constitutes an experiment that is
$\sim 2$ times more sensitive than those reported in the BATSE
catalogs (\cite{Batse1b}; \cite{Batse3b}; \cite{Batse4b};
\cite{Batse5b}).

In this paper we present results regarding the peak flux distribution
of the GRBs detected with our ``off-line'' search of archival data.
In section \ref{sec:search} we summarize some important aspects of our
off-line search and we discuss the \vvmax\ statistic for the bursts we
detected.  We show that surprisingly few bursts are found on the 4.096
s and 8.192 s time scales that were not already detected on the 1.024
s time scale.  In section \ref{sec:cosmo} we fit parametric
cosmological models to the observed differential peak flux
distribution to compare scenarios in which the GRB rate follows the
SFR with the model in which the co-moving GRB rate is independent of
redshift.  We also examine the possibility that a homogeneous (in
Euclidean space) population of bursting objects could be contributing
to the observed sample of GRBs.  In section \ref{sec:discuss} we show
how our results provide two independent arguments that favor models in
which the GRB rate follows the SFR over models in which the GRB rate
is independent of redshift.

\section{The Search for Non-Triggered GRBs}
\label{sec:search}
The details of our off-line search of the BATSE data are discussed in
Kommers et al.\ (1997).  We have merely extended the search from
covering 345 days of the mission to covering 2200 days.  We have also
made minor modifications to our peak flux estimation procedure in
order to secure better relative calibration between our peak fluxes
and those in the 4B catalog (\cite{Batse4b}).  The extended catalog of
non-triggered events will be provided and discussed in the
Non-Triggered Supplement to the BATSE Gamma-Ray Burst Catalogs
(\cite{Kommers99}).  Here we address only those aspects of the search
that are relevant to the GRB intensity distribution analysis.

We use the data from the Large Area Detectors that provide count rates
in 4 energy channels with 1.024 s time resolution (the data type
designated ``DISCLA'' in the flight software; \cite{Fishman89}).
These data are searched for statistically significant count rate
increases to identify candidate burst events.  The many candidate
events (``off-line triggers'') are then visually inspected to separate
astronomically interesting transients from instrumental and
terrestrial effects.  To be considered a GRB, a candidate must exhibit
significant signal in the 50--300 keV range (DISCLA channels 2 and 3)
and it must {\it lack\/} any characteristics that would associate it
with a solar flare, Earth magnetospheric particle precipitation, or
other non-GRB origin.  Since the DISCLA data are (nearly) continuously
recorded, our search detects some bursts that already activated the
onboard burst trigger; we call these events ``onboard-triggered
bursts.''  Bursts that were detected {\it exclusively\/} by our search
of archival data are called ``non-triggered bursts.''

In addition to searching at the 1.024 s time resolution of the DISCLA
files, we also search the data binned at 4.096 s and 8.192 s time
resolution.  The longer time bins provide greater sensitivity to faint
bursts that have durations longer than $\sim 4$ s or $\sim 8$ s.  The
specific time profile of each burst determines which of these three
time scales is the most sensitive.  For this reason the searches on
each time scale should be considered separate experiments.

Our search covers $1.33 \times 10^{8}$ s of archival data
spanning the time from 1991 December 9 to 1997 December 16.  In these
data we detected 2265 GRBs, of which 1392 activated the onboard burst
trigger and 873 did not.  We will refer to these 2265 GRBs as the
``off-line GRB sample.''  During the same time period, the onboard
burst trigger detected 1815 GRBs.  The $1815 - 1392 = 423$ bursts that
were detected by the onboard burst trigger but that were {\it not\/}
detected by our search either occurred during gaps in the archival
DISCLA data or had durations much less than the 1.024 s time
resolution (so they did not achieve adequate statistical significance
in the archival data).

{\it Note that because the best time resolution available to our
  retrospective search is 1.024 s, all results in this paper pertain
  to bursts with durations longer than about 1 s}.  Thus, the
population of ``short'' (duration less than $\sim 2$ s) bursts that
contributes to the bi-modal GRB duration distribution
(\cite{Kouveliotou93}) is not well represented in the off-line sample.
An estimate for the fraction of bursts that our search is likely to
miss because of our time resolution can be obtained from the 4B
catalog.  Although 21\% of GRBs for which both durations and fluences
were available had $T_{90} < 1.024$ s, only 7\% had both $T_{90} <
1.024$ s and fluences too small to create adequate statistical
significance in the 1.024 s data (\cite{Batse4b}).

For each of the 873 non-triggered GRBs we have estimated a peak flux
in the 50--300 keV range based on the time bin with the most counts
above background.  For 1288 of the 1392 onboard-triggered GRBs, we
used the peak fluxes from the current BATSE GRB catalog
(\cite{Batse5b}).  For the remaining 104 onboard-triggered bursts,
peak fluxes were not available from the current burst catalog; we
estimated peak fluxes for them using our own techniques as we did for
the non-triggered bursts.

Since the onboard trigger criteria were changed for a variety of
reasons during the time spanned by our search, we adopt for the
nominal onboard detection threshold the value 0.3 ph cm$^{-2}$
s$^{-1}$ in the 50--300 keV range.  At this peak flux the onboard
trigger efficiency is $\approx 0.5$ (\cite{Batse4b}).  With this
estimate, 551 of our 873 non-triggered bursts were below the nominal
onboard detection threshold.  The rest were not detected onboard for
the reasons cited previously.

\subsection{Trigger Efficiency}
\label{subsec:teff}
\begin{figure}
\plotone{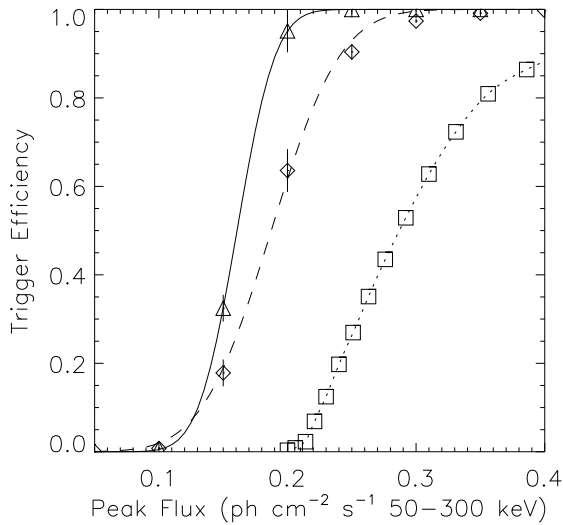}
\caption{Trigger efficiency for our off-line search.  The grid points
  of the calculations are plotted as individual symbols.  Error bars
  represent the standard deviations of the calculated probabilities
  owing to variations in the background rates.  The dashed line
  (equation 1) shows the probability that a burst occupying a single
  time bin is detected by our search.  The solid line (equation 4)
  shows the marginal probability that a burst is detected by our
  search, given that some bursts longer than 1.024 s have more than
  one statistical chance to be detected.  For comparison, the dotted
  line shows the trigger efficiency from the 4B catalog; no
  uncertainties are available for the grid points (squares).}
\label{fig:nteff}
\end{figure}

To determine the peak flux threshold of the off-line GRB sample, the
trigger efficiency $E_1(P)$ of our off-line search has been calculated
using the techniques described in Kommers et al.\ (1997).  This
quantity is the probability that a burst that occupies exactly one
1.024 s time bin with a peak flux $P$ will be detected by the off-line
search algorithm.  $E_1(P)$ is well represented (within the
uncertainties of the calculation owing to variations in the background
rates) by the following function:
\begin{equation}
\label{eqn:oneteff}
E_1(P) = \frac{1}{2} \left[ 1 + {\rm erf}(-3.125 + 16.677 P) \right],
\end{equation}
where ${\rm erf}(x)$ is the standard error function and $P$ is given
in units of photons cm$^{-2}$ s$^{-1}$ in the 50-300 keV band.  This
equation is plotted as the dashed line in Figure~\ref{fig:nteff}.
Error bars on the grid points of the calculation (diamonds) represent
the sample standard deviation of the calculated probabilities owing to
variations in the background rates.  For comparison, the BATSE trigger
efficiency from the 4B catalog (\cite{Batse4b}) is plotted as the
dotted line (grid points are indicated by open squares).  Equation
\ref{eqn:oneteff} tends to {\it underestimate\/} the probability that
a typical GRB will be detected, however.  This is because many GRBs in
our sample last longer than 1.024 s; therefore, these bursts have more
than one statistical chance to be included in the sample.

\begin{figure}
\plotone{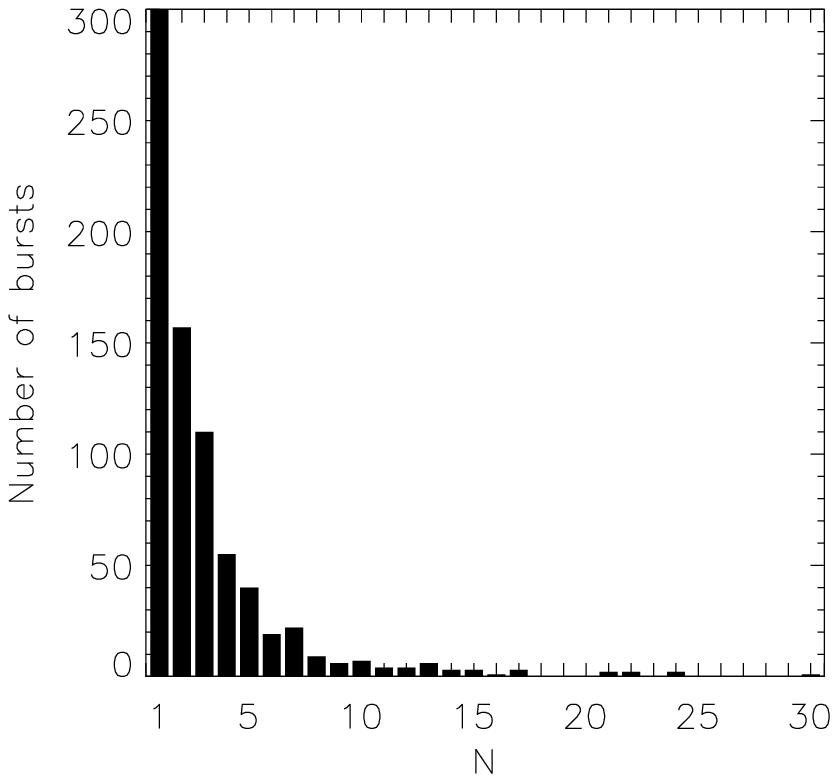}
\caption{Histogram of $N$, the number of time bins within one standard
  deviation of the peak count rate, for bursts with peak fluxes in the
  range 0.1--0.4 ph \protect{cm$^{-2}$ s$^{-1}$}.}
\label{fig:nhist}
\end{figure}

Suppose the peak of a burst occupies $N$ time bins, so that the burst
has effectively $N$ statistical chances to be detected. Then the
probability that the burst is detected can be approximated as unity
minus the probability that the burst {\it fails\/} to be detected in
all $N$ trials:
\begin{equation}
E_N(P) = 1 - \left[ 1 - E_1(P) \right]^{N}.
\end{equation}
Since the number of chances $N$ is not known for GRBs {\it a priori},
the actual probability of detection $E(P)$ is obtained by
marginalizing $E_N(P)$ over the distribution of $N$ for bursts with
peak fluxes near $P$:
\begin{equation}
E(P) = \frac{\sum h(N,P)\,E_N(P)}{\sum h(N,P)}.
\end{equation} 
Our estimate for $h(N,P)$, the histogram of the various integer values
of $N$ for bursts with peak fluxes near $P$, was obtained from the
detected sample of bursts by counting, for each burst, the number of
time bins with count rates that were within one standard deviation of
the peak count rate.  For purposes of illustration,
Figure~\ref{fig:nhist} shows the histogram of $N$ for bursts with peak
fluxes in the range 0.1--0.4 ph cm$^{-2}$ s$^{-1}$.  The resulting
function $E(P)$ is well represented (to within the uncertainties of
the calculation) by the formula
\begin{equation}
\label{eqn:nteff}
E(P) = \frac{1}{2} \left[ 1 + {\rm erf}(-4.801 + 29.868 P) \right].
\end{equation}
This equation expresses our best estimate of the trigger efficiency of
our off-line search on the 1.024 s time scale.  It is plotted as the
solid line in Figure~\ref{fig:nteff}.  The efficiency of our search
falls below 0.5 at a peak flux of 0.16 ph cm$^{-2}$ s$^{-1}$.

If we had not made some correction for the effect of time profiles on
the single-time-bin burst detection probabilities, we would have
substantially underestimated our trigger efficiency near the detection
threshold ($\sim 0.2$ ph cm$^{-2}$ s$^{-1}$).  We note that this type
of correction to the single-time-bin trigger efficiency should also be
applied when using the trigger efficiencies given in the 1B, 2B, 3B,
and 4B catalogs (\cite{Batse1b}; \cite{Batse3b}; \cite{Batse4b}).
Similar considerations are addressed by in't Zand \& Fenimore (1994)
and Loredo \& Wasserman (1995).

\subsection{$(C_{min}/C_{max})^{3/2}$ Distribution}
\label{subsec:vvmax}
As successively more sensitive instruments have been used to produce
GRB catalogs, it has been customary to give the value of the \vvmax\ 
statistic for the detected bursts (\cite{Schmidt88}).  For photon
counting experiments like BATSE, it is not strictly \vvmax\ that is
typically calculated, but rather $\langle (C_{min}/C_{max})^{3/2}
\rangle$, where $C_{min}$ is the threshold count rate and $C_{max}$ is
the maximum count rate measured during the burst.  The departure of
$\langle (C_{min}/C_{max})^{3/2} \rangle$ from the value of
$\frac{1}{2}$ expected for a population of bursters distributed
homogeneously in Euclidean space (with a well-behaved, but otherwise
arbitrary luminosity distribution) has been firmly established
(\cite{Meegan92}; \cite{Batse3b}).  Since the discovery that most GRBs
originate at cosmological distances, the quantity $\langle
(C_{min}/C_{max})^{3/2} \rangle$ can no longer be interpreted as
\vvmax.  Nevertheless, it is useful to compare the values of $\langle
(C_{min}/C_{max})^{3/2} \rangle$ obtained by successively more
sensitive experiments, including the value obtained for the bursts
detected with our search.

Table~\ref{tbl:vvmax} lists various missions and the values they
obtained for $\langle (C_{min}/C_{max})^{3/2} \rangle$.  The trend
towards lower values of $\langle (C_{min}/C_{max})^{3/2} \rangle$ with
more sensitive experiments indicates that increasing the accessible
survey volume by decreasing the flux threshold does not lead to the
detection of large numbers of faint bursts.

\begin{figure}
\plotone{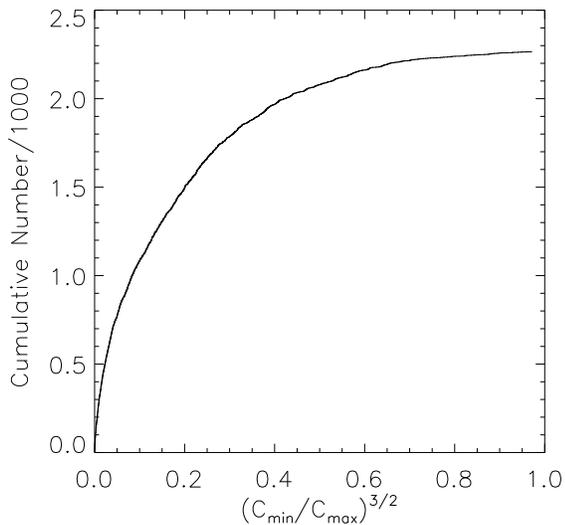}
\caption{Cumulative distribution of
  \protect{$(C_{min}/C_{max})^{3/2}$} for the off-line GRB sample.
  The dramatic flattening of the curve above
  \protect{$(C_{min}/C_{max})^{3/2} = 0.5$} shows that few of the
  GRBs detected in our search are just barely above the detection
  threshold on all 3 time scales (1.024 s, 4.096 s, and 8.192 s).}
\label{fig:vvmax}
\end{figure}

The value of $\langle (C_{min}/C_{max})^{3/2} \rangle$ for the 2265
GRBs detected by our search\footnote{This value supersedes the ones
  given in Kommers et al.\ (1996, 1997, 1998), which are incorrect
  due to a programming error.  An erratum has been submitted (Kommers
  et al.\ 1999a).} is $0.177 \pm 0.006$.  {\it This is the lowest
  value ever obtained for a sample of GRBs}.  The cumulative
distribution of $(C_{min}/C_{max})^{3/2}$ for our GRBs is shown in
Figure~\ref{fig:vvmax}.  The flattening of this curve in the range
$0.5 < (C_{min}/C_{max})^{3/2} < 1.0$ shows that over 90\% of the GRBs
we detect are above threshold (on at least one of the 3 time scales)
by a factor of at least $(0.5)^{-3/2} = 1.6$.

The reason for this low value of $\langle (C_{min}/C_{max})^{3/2}
\rangle$ is the fact that most of the bursts we detected had their
maximum signal-to-noise ratios on the 4.096 s and 8.192 s time scales,
yet surprisingly few bursts were detected {\it only\/} on these
longer time scales.  For each burst we compute the values of
$(C_{min}/C_{max})^{3/2}$ on each of the 3 time scales.  The largest
of the three values for each burst is used in taking the average.  In
Euclidean space this corresponds to taking for each burst the {\it
  smallest\/} value of \vvmax.  Since 72.0\% of the bursts we detected
have $T_{90}$ durations (\cite{Koshut96}) longer than 8 s, we expect
the average $\langle (C_{min}/C_{max})^{3/2} \rangle$ to be dominated
by values measured on the 8.192 s time scale.

In fact, the average $\langle (C_{min}/C_{max})^{3/2} \rangle = 0.177
\pm 0.006$ includes 520 values measured on the 1.024 s time scale, 491
values measured on the 4.096 s time scale, and 1254 values measured on
the 8.192 s time scale.  Yet only 105 bursts were detected {\it
  exclusively \/} on either of the 4.096 or 8.192 s time scales (or
both).  Many of the bursts that are barely above the detection
threshold on the 1.024 s time scale are well above the detection
threshold on the longer time scales.  Thus very few bursts are found
to be just barely above our detection threshold on all 3 time scales,
and this accounts for the low value of $\langle
(C_{min}/C_{max})^{3/2} \rangle$.  Restricting our calculation to use
{\it only\/} count rates measured on the 1.024 s time scale (and
bursts detected on the 1.024 s time scale) gives a larger value,
$\langle (C_{min}/C_{max})^{3/2} \rangle = 0.247 \pm 0.006$.

Roughly, the 4.096 s search should be $\sim 2$ times more sensitive
than the 1.024 s search for bursts that maintain their peak flux for
at least $\sim 4$ s, and the 8.192 s search should be yet more
sensitive.  Therefore our lack of GRB detections exclusively on the
longer time scales indicates either 1) a substantial paucity of faint,
long bursts below the threshold of our 1.024 s search, or 2) that
during our visual inspection of the off-line triggers we have tended
to classify a substantial number of faint, long GRBs as other
(non-GRB) phenomena.  We feel that both alternatives must be present
at some level.  

A review of the non-GRB off-line triggers suggests that events
resembling faint, long GRBs that illuminate the same detectors as a
known, bright, variable X-ray source are more likely to be attributed
to variability from the X-ray source than to be classified as GRBs.
There is also a tendency to classify bursts that have directions
consistent with the Sun as solar flares.  A secondary evaluation of
the event classifications suggests that between 50 and 200 (this range
represents the central 90\% confidence interval), with a most likely
value of 86, GRBs have been misclassified in this way.  The
corresponding ``loss rate'' is between 2\% and 8\% (most likely 4\%)
of the total 2265 bursts in the off-line sample.  This is not enough
to fully explain, as experimental error, the dearth of faint, long
bursts below our 1.024 s threshold.

\subsection{Peak Fluxes}
\label{subsec:pfdata}
Detailed comparisons of cosmological models with the data require
intensity distributions in physical units.  We have chosen to do the
analysis in terms of the burst rate as a function of peak photon flux
measured over 1.024 s in the energy range 50--300 keV.  Compared with
the fluence (total energy per unit area deposited in the detector by
the burst) we prefer peak photon flux for the purposes of intensity
analysis.  The peak photon flux can be obtained more reliably
from the raw count data and it is more directly related to our ability
to detect bursts.

Of the 2265 GRBs detected by our search, we chose to include in our
peak flux analysis only those that were detected on the 1.024 s time
scale, so that equation \ref{eqn:nteff} gives the detection
efficiency.  We also chose to use only those bursts with peak fluxes
in the range 0.18--20.0 ph cm$^{-2}$ s$^{-1}$.  The lower limit
ensures that the off-line trigger efficiency exceeds 0.8 for the range of
intensities used in the analysis, and the upper limit excludes very
bright bursts which are too rare to provide adequate counting
statistics in narrow peak flux bins.  With these cuts on the data, we
are left with 1998 peak flux measurements.  To fit the differential
intensity distribution, we bin the 1998 bursts into 25 peak flux
intervals that were chosen to be approximately evenly spaced in the
logarithm of $P$.  The spacing is $\Delta \log P \approx 0.05$ in the
range $0.18 < P < 1.0$, $\Delta \log P \approx 0.1$ in the range $1.0
< P < 7.9$, and there is a final broad bin for the range $7.9 < P <
20.0$.  Uncertainties in the number of bursts $\Delta N_{obs}$ in each
bin are taken to be $\pm \sqrt{\Delta N_{obs}}$.  The burst rate is
computed by dividing the number of bursts in each bin by the live time
of the search ($1.33 \times 10^{8}$ s = 4.21 yr) and the mean solid
angle visible to the BATSE detectors ($0.67 \times 4 \pi$).
Table~\ref{tbl:pfdata} gives the peak flux intervals, number of
bursts, and burst rate for each bin.

\section{Cosmological Model Comparison}
\label{sec:cosmo}
Many investigators, in scores of papers, have shown the consistency of
the GRB peak flux distribution with various cosmological models (see,
for example, \cite{Wijers98}; \cite{Loredo98}; \cite{Hakkila96};
\cite{Horack96}; \cite{Rutledge95}; \cite{Fenimore95}; and references
therein).  As shown in the previous section, the off-line GRB sample
extends the observed GRB intensity distribution to peak fluxes that
are lower by a factor of $\sim 2$ than could be studied previously.
While it is unlikely that a factor of $\sim 2$ will yield stringent
new model constraints, it remains of interest to note a few
cosmological models that provide good fits to the extended GRB peak
flux distribution.  These can be used to set limits on the rate of
GRBs that may come from a nearby, spatially homogeneous subpopulation of burst
sources.

\subsection{Purely Cosmological Models}
\label{subsec:purecos}
To limit the number of free parameters that must be considered, our
choice of cosmological world model is the Einstein-de Sitter model
($\Omega = 1$, $\Lambda = 0$, $q_0 = \frac{1}{2}$; \cite{Weinberg72}).
This cosmology has been used by many other investigators so it allows
easy comparison of results.  Where needed, we assume a Hubble constant
$H_0 = 70\ h_{70}$ km s$^{-1}$ Mpc$^{-1}$.  We also assume that
bursters are distributed isotropically, so the only interesting
parameter in the burster spatial (redshift) distribution is the radial
coordinate $r(z)$ from Earth.  The following derivation of the
expected peak flux distributions follows the discussions in
Fenimore \& Bloom (1995) and Loredo \& Wasserman (1997).

In general the rate of bursts $R$ per unit interval in peak flux $P$
observable at Earth is given by
\begin{equation}
\label{eqn:genrate}
\frac{dR}{dP} =  \int\!dL\!\int\!dz\ \frac{\partial^{2}R}{\partial L\,
  \partial z}\ \delta(P -
\Phi(L,z)),
\end{equation}
where $L$ is the equivalent isotropic peak luminosity of the burst at
the source, $z$ is the redshift parameter, $\partial^{2} R/\partial
L\, \partial z$ is the rate
of bursts per unit $L$ per unit redshift interval, $\delta(x)$ is the
Dirac delta function, and $\Phi(L,z)$ is the peak photon flux measured
at Earth for a burst with peak luminosity $L$ located at redshift $z$.
We will assume that the redshift and luminosity distributions are
independent, so that the burst rate as a function of $L$ and $z$ is
given by
\begin{equation}
\frac{\partial^2 R}{\partial L\, \partial z} = \frac{4 \pi c R_0}{H_0}\ \psi(L)\rho(z)\ 
\frac{r^2(z)}{(1+z)^2\ \sqrt{1 + z}}\ 
\end{equation}
where $R_0$ is an overall normalization, $\psi(L)$ is the distribution
of burst luminosities (normalized to unity), $\rho(z)$ is the
distribution of the co-moving burst rate as a function of redshift
(normalized to unity on the interval $0 < z < 10$), and $r(z) =
(2c/H_0) (1 + z - \sqrt{1 + z})/(1+z)$ is the co-moving radial
coordinate.

The peak flux $\Phi(L,z)$ observed at Earth in the 50--300 keV energy
range, where the BATSE burst trigger is sensitive, depends on the
intrinsic spectrum of the GRB.  We write it as
\begin{equation}
\Phi(L,z) = \frac{L\ K(z)}{4 \pi\ (1+z)\ r^2(z)}.
\end{equation}
The spectral correction function $K(z)$ depends on the shape of the
burst photon energy spectrum at the source.  The observed GRBs have a
variety of spectral shapes, and in the cosmological scenario these
observed spectra have been redshifted according to the (unknown)
redshifts of the sources.  

To account for the spectral variety of GRBs we use the spectral fits
of Band et al.\ (1993).  To account for the unknown redshift factors
for these spectra, we use the procedure described in Fenimore \& Bloom
(1995).  The peak fluxes of the bursts for which Band et al.\ (1993)
derived spectral fits are used in conjunction with the cosmological
model under consideration to self-consistently estimate the redshift
factors for the fitted spectra.  We assume that the $i$th burst fitted
by Band et al.\ (1993) has exactly the mean intrinsic peak luminosity
in the cosmological model being considered: $L_i = \int\!dL'\ L'
\psi(L')$, where the shape of $\psi(L)$ depends on the parameters of
the model luminosity function.  We then solve for the redshift $z_i$
which the fitted burst $i$ must have had to produce the peak flux
listed for it in the current BATSE GRB catalog (\cite{Batse5b}).
Fifty-one of the bursts fitted by Band et al.\ (1993) had peak fluxes
available.  For each of their spectral shapes $\phi_i(E)$ the spectral
correction function takes the form
\begin{equation}
K_i(z) = \frac{\int^{300(1 +
    z)/(1+z_i)}_{50(1+z)/(1+z_i)}\!dE\
    \phi_i(E)}{(1+z_i)\ \int^{2000/(1+z_i)}_{30/(1+z_i)}\!dE\ E \phi_i(E)}.
\end{equation}
The integrals in the denominator and numerator convert the model
parameter $L$, which represents the peak luminosity in the 30--2000
keV range at the source, to the observed photons cm$^{-2}$ s$^{-1}$ in
the 50-300 keV band at Earth. The burst rate expected in the BATSE
band pass for the $i$th spectral shape $\phi_i(E)$ is then (from
equation \ref{eqn:genrate})
\small
\begin{eqnarray}
\label{eqn:rateint}
\lefteqn{\left( \frac{dR}{dP} \right)_i = \frac{16 \pi^2 c R_0}{H_0}\ \times }
\\ & \int\!dz\
\frac{\rho(z)\ r^4(z)}{(1+z)\ \sqrt{1 + z}\ K_i(z)}\ \psi \left( \frac{4 \pi\
    (1+z)\ r^2(z)\ P}{K_i(z)} \right) \nonumber.
\end{eqnarray}
\normalsize
The limits on the integral are determined by the range of $z$ for
which $\psi(4 \pi (1+z) r^2(z) P/K_i(z))$ is non-zero at the given
$P$.
 
To estimate the observed distribution of bursts, which includes a
variety of spectral shapes, we average Equation~\ref{eqn:rateint} over
the 51 spectral correction functions $K_i(z)$.  This procedure is
equivalent to marginalizing the unknown spectral parameters of the
observed bursts (i.e., those in the off-line sample) to obtain the
posterior rate distribution. The 51 spectra from Band et al.\ (1993)
are furnishing estimates of the prior distributions of the spectral
parameters.  The expectation value of the observed burst rate for peak
fluxes between $P_1$ and $P_2$ is then
\begin{eqnarray}
\label{eqn:rate}
\lefteqn{\Delta R(P_1, P_2) = } \\ & \int_{P_1}^{P_2}\!dP\ E(P)\ \left\langle \frac{dR}{dP}
\right\rangle \nonumber,
\end{eqnarray}
where $\langle dR/dP \rangle$ is the mean rate estimated from the 51
observed spectra and $E(P)$ is the detection efficiency.

The use of the Band et al.\ spectra increases the computational cost
of the rate model by a factor of $\sim 50$ over using a single
``universal'' burst spectrum.  We found that a simple power-law form
for the GRB photon energy spectrum---as has been used by many previous
studies---predicts significantly different burst rates at low peak
fluxes than does equation \ref{eqn:rate}.  Since we are interested in
the behavior of the burst rate at low peak fluxes, we felt that the
analysis based on the full 51 Band et al.\ spectra would be more
reliable.  Similar conclusions are reached by Fenimore \& Bloom (1995)
and Mallozzi, Pendleton, \& Paciesas (1996).

For comparison with the results of previous studies, we chose two
forms for the luminosity distribution.  The first is a monoluminous
(standard candle) distribution.  The second is a truncated power-law,
\begin{equation}
\psi(L) = \left\{ \begin{array}{ll} \frac{1}{L
      \log(L_{max}/L_{min})} & \beta = 1 \\
      \frac{(1-\beta)\ L^{-\beta}}{L_{min}^{1 - \beta} -
      L_{max}^{1 - \beta}} & \beta \neq 1 \end{array} \right.
\end{equation}
with $\psi(L) = 0$ if $L < L_{min}$ or $ L > L_{max}$.  The
normalization factors ensure that $\int\!dL\ \psi(L) = 1$.  

The standard candle distribution, though useful for comparison with
other results, is ruled out by the observed peak fluxes of the four
bursts for which associated optical redshifts have been measured.  For
GRBs 970508, 971214, 980613, and 980703 the inferred equivalent
isotropic peak luminosities in the 30--2000 keV energy range are given
in Table \ref{tbl:rlums}.  To calculate each of these peak
luminosities, we have used the observed 50--300 keV peak flux (on the
1.024 s time scale) in combination with the observed redshift to find
the expectation value of the intrinsic luminosity averaged over the 51
Band et al.\ spectra.  This procedure is the one used in our modeling,
so it was used on these four bursts also, to facilitate comparisons
with the models (see section \ref{sec:discuss}).  The peak
luminosities estimated here are somewhat higher by factors of $\sim 3$
to $\sim 6$ than those reported elsewhere (e.g., Krumholz et al.\ 
1998).  This is because the spectral shapes fitted by Band et al.\ 
(1993) generally become steeper at high energies, so a source at high
redshift must be more luminous to produce the flux observed at Earth
than it would have to be if the spectrum did not fall off so rapidly
at higher energies.  These differences illustrate the importance of
using the most realistic spectral models available rather than simple
power-laws when analyzing the GRB intensity distribution.

A variety of spatial, or rather redshift, distributions for the
bursters have been used in previous studies of the GRB intensity
distribution.  With up to 4 free parameters already incorporated into
our burst rate models (the overall normalization $R_0$, and the
parameters of the power-law luminosity function $\beta$, $L_{min}$,
and $L_{max}$) there is little hope of constraining any additional
free parameters in the redshift distribution.  Here we explore 3
specific models of the redshift distribution that contain no free
parameters.  The two physical scenarios we examine are 1) that the
co-moving burst rate is independent of redshift between $z=0$ and
$z=10$, and 2) that the co-moving GRB rate is proportional to the star
formation rate (SFR).

For the GRB rate model that is independent of redshift, $\rho(z) =
0.1$ for $0 < z \leq 10$ and $\rho(z) = 0$ for $z > 10$.  We refer to
this redshift distribution as ``model D1.''

For the case where the burst rate follows the star formation history
of the Universe, we use two slightly different parameterizations of
the SFR.  The first is the SFR deduced from the rest-frame ultraviolet
luminosity density, with the functional form given in footnote 1 of
Madau, Della Valle, \& Panagia (1998).  In this estimation the SFR
peaks around $z = $1--1.5.  A SFR of roughly this form has been
used by several previous studies of the GRB intensity distribution
(\cite{Totani97}; \cite{Wijers98}; Krumholz et al.\ 1998).  We refer to this
redshift distribution as ``model D2.''

The second SFR parameterization assumes that the SFR---and thus the
GRB rate---tracks the total output of radio-loud AGN.  In this
scenario the SFR peaks at $z=$2--3 (\cite{Hughes98}; \cite{Dunlop98}).
This form of the SFR appears to be more consistent with recent results
from SCUBA (\cite{Hughes98}) which are not susceptible to the same
problems of dust obscuration as the determination by Madau et al.\ 
(1998).  The specific functional form we use is a best-fit analytic
model to points measured by hand from Figure~6 of Hughes et al.\ (1998):
$\rho(z) \propto 0.00360 + 0.0108 \exp (2.76 z - 0.573z^2)$.  This
approximation appears to be accurate to within 5\% for the redshift
range $1 < z < 4$.  (At lower and higher redshifts the formula likely
underestimates the actual rate of star formation; but this is no great
concern as it is the redshift of the peak SFR that is of primary
interest.)  We refer to this redshift distribution as ``model D3.''

With choices for $\psi(L)$ and $\rho(z)$ as discussed, we fit equation
\ref{eqn:rate} to the data in Table~\ref{tbl:pfdata} by minimizing the
$\chi^2$ statistic.  In all cases, we found that the parameter
$L_{max}$ was not well constrained: variations in $L_{max}$ did not
change the minimum $\chi^2$ by a significant amount.  The
(mathematical) reason for this is that the integrand in equation
\ref{eqn:rateint} is a decreasing function of $z$ for plausible values
of $\beta$, so that varying the upper limit ($z_{max}$ corresponding
to $L_{max}$ for the given $P$) causes only small changes in the value
of the integral.  Accordingly, all the results reported here set
$L_{max} \equiv 1000L_{min}$.  The free parameters are thus $R_0$ and
$L_{0}$ in the cases of the standard candle models, and $R_0$,
$\beta$, and $L_{min}$ in the cases of the power-law luminosity
distribution models.  The results of the fits are listed in
Table~\ref{tbl:fitresstd} and Table~\ref{tbl:fitrespl}.  Uncertainties
on the fitted parameters correspond to 68\% confidence limits for 2
($\Delta \chi^2 = 2.3$) or 3 ($\Delta \chi^2 = 3.5$) interesting
parameters, respectively (\cite{Avni76}).

Model D1 (constant burst rate density as a function of redshift)
produces an acceptable fit for the standard candle luminosity
distribution.  The probability of getting $\chi^2 > 32.3$ for 23
degrees of freedom is 0.094.  Adding one more free parameter ($\beta$)
for the power-law luminosity distribution produces an insignificant
change in the minimum $\chi^2$.  Furthermore, the high value of
$\beta$ in the best-fit power-law distribution indicates a very narrow
range of peak luminosities.

Model D2 (burst rate density follows the SFR as determined by Madau et
al.\ 1998) produces a formally unacceptable fit in the monoluminous
case.  But it achieves an excellent fit ($\chi^2$ per degree of
freedom = 0.81) for the power-law luminosity distribution.  The
$F$-test estimates a probability of $1.5 \times 10^{-7}$ that the
improvement in $\chi^2$ is due to chance, justifying the inclusion of
the additional parameter in the power-law luminosity function model.
The value of $\beta$ in this model is remarkably well-constrained.  If
the other fit parameters are regarded as ``uninteresting'' then the
90\% ($\Delta \chi^2$ = 2.7) and 99\% ($\Delta \chi^2$ = 6.6)
confidence intervals (\cite{Avni76}) on $\beta$ are 2.0--2.3 and
1.8--2.6, respectively.

Model D3 (burst rate density follows the output of radio-loud AGN)
produces formally acceptable fits with both the standard candle and
power-law luminosity distributions.  The power-law luminosity
distribution achieves a significantly lower $\chi^2$, however.  The
$F$-test estimate of the probability that the improvement is due to
chance is $1.5 \times 10^{-3}$.

\begin{figure}
\plotone{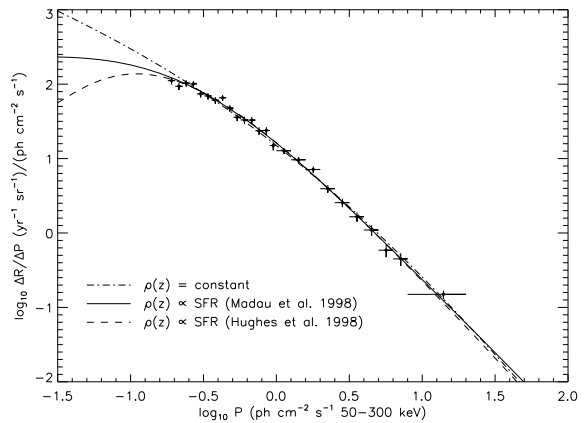}
\caption{Best-fit cosmological models with power-law luminosity
  distributions.  Units of \protect{$R$} are bursts \protect{yr$^{-1}$
    sr$^{-1}$}, and those of \protect{$P$} are \protect{ph cm$^{-2}$
    s$^{-1}$} in 50--300 keV.  The dot-dashed line corresponds to
  model D1 (co-moving burst rate is independent of redshift).  The
  solid line shows model D2 (burst rate follows the rest-frame
  ultraviolet luminosity density) and the dashed line shows model D3
  (burst rate follows the output of radio-loud AGN).  Measured rates
  are shown with 1$\sigma$ vertical error bars; horizontal error bars
  indicate the bin widths.  The best-fit model curves displayed here
  have not been corrected for detection efficiency.}
\label{fig:bestfit}
\end{figure}

Figure~\ref{fig:bestfit} plots the differential peak flux
distributions for the best-fit models with power-law luminosity
distributions.  For all three best-fit models the value of $\langle
(P_{min}/P)^{3/2} \rangle$ is consistent with the value of $\langle
(C_{min}/C_{max})^{3/2} \rangle$ measured for the sample (see section
\ref{subsec:vvmax}).  Extrapolating the best-fit models to peak fluxes
lower than those included in our data shows very different behaviors.
Model D1 (dot-dashed lines) predicts a dramatically higher burst rate
at low peak fluxes than do models D2 (solid line) and D3 (dashed
line).

In each model the best-fit parameters for the power-law luminosity
function yield our best estimate of the parameters of the {\it
  intrinsic\/} distribution of GRB peak luminosities.  The
distribution of peak luminosities of the {\it observed\/} bursts is
different, however, because the most luminous bursts are sampled from
a much larger volume than than are the least luminous bursts.  Even
though high luminosity bursts are infrequent, the geometrical
advantage of sampling them from a larger volume means that they will
be over-represented in a sample of bursts observed over a fixed time
interval.  The distribution of peak luminosities for the observed
bursts is the ``effective luminosity function'' (see \cite{Loredo97}
for further discussion).  For the best-fit parameters of model D1 the
effective luminosity function is a power-law that is less steep than
that of the intrinsic luminosity function.  We find $\beta_{{\rm
    eff}}^{{\rm D1}} = 2.8$ for the effective luminosity function
versus $\beta = 4.6$ for the intrinsic one.  The power-law slopes of
the effective luminosity functions for models D2 and D3 are
$\beta_{{\rm eff}}^{{\rm D2}} = 1.6$ and $\beta_{{\rm eff}}^{{\rm D3}}
= 1.9$, respectively.

\begin{figure}
\plotone{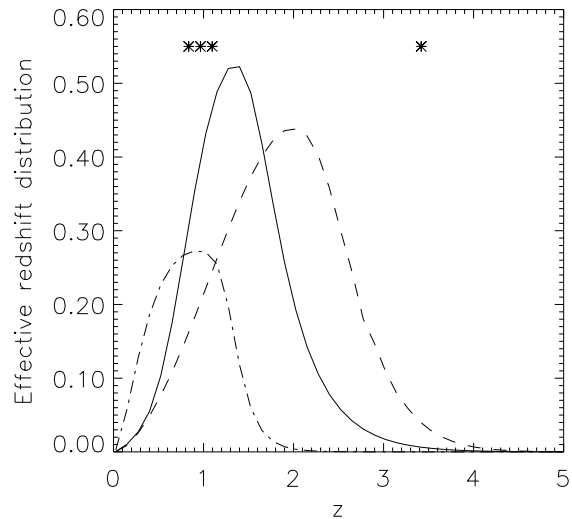}
\caption{Redshift distribution of the burst rate for observed bursts
  (i.e., the effective differential burst rate as function of
  redshift).  The distributions are normalized so that the integral of
  the distribution visible to a ``perfect'' detector with no
  sensitivity limit is unity.  The dot-dashed line corresponds to a
  constant burst rate as a function of redshift (model D1).  The solid
  line corresponds to the model (D2) where the burst rate traces the
  SFR as determined by Madau et al.\ (1998).  The dashed line
  corresponds to the model (D3) where the burst rate traces the output
  of radio-loud AGN (Hughes et al.\ 1998).  In model D1 the off-line
  search detects a smaller percentage of the bursts that occur than it
  does in models D2 and D3.  The redshifts associated with GRBs
  970508, 971214, 980613, and 980703 are marked with asterisks.}
\label{fig:effzee}
\end{figure}

Similarly, the distribution of the GRB rate as a function of redshift
for the observed bursts is not identical to the intrinsic redshift
distribution given by $\rho(z)$ (see \cite{Loredo97}).
Figure~\ref{fig:effzee} shows the effective redshift distributions for
the best-fit models D1, D2, and D3.  In all 3 models the effective
redshift distribution cuts off at a lower redshift than does the
corresponding intrinsic redshift distribution.  The mean redshifts of
the observed bursts in the best-fit models D1, D2, and D3 are $\langle
z \rangle^{D1}$ = 0.86, $\langle z \rangle^{D2}$ = 1.4, and $\langle z
\rangle^{D3}$ = 1.9, respectively.  The maximum redshifts of the
bursts in models D1, D2, and D3 are not precisely determined, but the
observed redshift distributions cut off around $z_{\rm max}^{\rm D1}
\approx 1.5$, $z_{\rm max}^{\rm D2} \approx 2$, and $z_{\rm max}^{\rm
  D3} \approx 3$, respectively (see Figure~\ref{fig:effzee}).

It is interesting to compare the rate of bursts that are seen with the
off-line search to the total rate of bursts that occur in the
Universe, subject to the cosmological rate models we are considering.
The fraction of bursts that are detected with BATSE is given by the
integral of the effective redshift distribution for the off-line
search (over the range $0 < z < 10$) divided by the integral of the
effective redshift distribution that would be visible to a ``perfect
detector'' which can detect infinitely faint bursts, multiplied by the
live-time (0.70) and sky exposure (0.67) fractions.  In the best-fit
models D1, D2, and D3, it follows that the off-line search detects
$\sim$ 14\%, $\sim$ 30\%, and $\sim$ 37\%, respectively, of the bursts
that occur in the Universe.  Of course, these estimates can apply only
to bursts with energy spectra and durations of the kind that are
accessible to the off-line search.  If BATSE had 100\% live-time and
no Earth blockage, then in model D1 30\% of the bursts that occur
could be detected with the off-line search.  In models D2 and D3, 65\%
and 78\%, respectively, of the bursts that occur could be detected
with the off-line search.

It is customary to quote the rate of GRBs as the co-moving rate per
unit volume at $z = 0$, a quantity often denoted by $\rho_0$ with
units of Gpc$^{-3}$ yr$^{-1}$ (\cite{Fenimore95,Wijers98}).  The model
parameter $R_0$ is related to $\rho_0$ by $\rho_0 = 4 \pi R_0 \rho
(0)$.  Table~\ref{tbl:ratestd} lists the values of $\rho_0$
corresponding to the best-fit values of $R_0$ for the standard-candle
GRB models.  Table~\ref{tbl:ratepl} does the same for the GRB models
with a power-law luminosity function.  This burst rate can be
converted into an event rate per ``typical'' galaxy using the space
density of such galaxies.  Here, a ``typical'' galaxy is taken to be
one with a luminosity $L^{\ast}$ equal to the characteristic
luminosity of galaxies in the Schechter function (\cite{Schechter76}).
Loveday et al.\ (1992) report the mean space density of
$L^{\ast}$ galaxies to be $(4.8 \pm 0.6)\ 10^{-3}\ h^3_{70}\ $
Mpc$^{-3}$.  With this conversion, the co-moving rate of GRBs at $z =
0$ can be expressed in GEM (Galactic Events per Million years), which
is the rate of GRBs in an $L^{\ast}$ galaxy.  This quantity is also
listed in Tables~\ref{tbl:ratestd} and \ref{tbl:ratepl}.

\subsection{Limits on a Possible Homogeneous Sub-population}
\label{subsec:subpop}
The discovery of the unusual Type Ib/c supernova SN 1998bw in the
X-ray error box of GRB 980425 has fueled speculation that the
supernova (SN) produced the GRB (\cite{Galama98}; \cite{Iwamoto98};
\cite{Kulkarni98b}).  It has been suggested that such events (the
supernova-GRBs, or ``S-GRBs'') may constitute a subclass of all GRBs
(\cite{Bloom98}).  In this subsection we discuss what fraction of all
GRBs could belong to such a subclass assuming that the remaining
bursts come from the ``reasonable'' cosmological scenarios discussed
previously.

If the inferred peak luminosity of GRB 980425 (assuming a distance
corresponding to the redshift $z = 0.0085$ of SN 1998bw;
\cite{Tinney98}) is typical of S-GRBs, then the bursts in this
subclass are detectable within a volume of radius $\sim$ 100 Mpc
(\cite{Bloom98}).  Within this volume, we assume the spatial
distribution of Type Ib/c SNe to be approximately homogeneous.  Thus
the cumulative intensity distribution of S-GRBs can be expected to
follow a $-3/2$ power-law.  This conclusion follows for any
well-behaved distribution of intrinsic luminosities as long the
spatial distribution does not deviate from homogeneity within the
volume sampled by our detectors.  Since the observed intensity
distribution of all GRBs deviates strongly from the $-3/2$ power-law,
it can be used to set an upper limit on the fraction of all GRBs that
can come from a subclass which obeys the $-3/2$ power-law.  In this
respect, the faint end of the peak flux distribution (as explored by
our off-line search) provides the most stringent constraints.

A model-independent limit on the fraction of bursts that might come
from a nearby homogeneous population can be obtained from the
histogram of $(C_{min}/C_{max})^{3/2}$ for the observed bursts.  Any
homogeneous subpopulation is expected to contribute a constant number
of bursts to each bin in the histogram.  Thus the bin with the fewest
number of bursts sets an upper limit on the total number of bursts
that could come from the subpopulation.  The bursts detected with our
search have already been used to set an upper limit of 5--6\% on the
fraction of bursts that could come from a homogeneous subpopulation;
this limit is therefore an upper limit on the fraction of bursts that
could be associated with nearby Type Ib/c SNe (\cite{Kippen98}).
Though model-independent, this limit depends on how coarsely the
histogram is binned.  It also assumes than an arbitrary distribution
of intensities is acceptable for bursts that do {\it not\/} come from
the homogeneous subpopulation.  This is too much freedom, because
physically plausible distributions occupy only a subset of all
arbitrary intensity distributions.

Here we assume that the bulk of GRBs come from the cosmological
distributions discussed in section \ref{subsec:purecos}.  An upper
limit on the rate of all GRBs that might come from Type Ib/c SNe (or
any nearby homogeneous distribution) is then fixed by determining the
maximum rate of bursts that can come from a (differential)
distribution proportional to $P^{-5/2}$ before the model becomes
inconsistent with the data.  We thus fit a model of the form
\small
\begin{eqnarray}
\Delta R(P_1, P_2) = & R_H\ \int_{P_1}^{P_2}\!dP\ E(P)\
  P^{-5/2}\ + \\ & \int_{P_1}^{P_2}\!dP\ E(P)\ \left\langle
  \frac{dR}{dP} \right\rangle \nonumber.
\end{eqnarray}
\normalsize
The fractional burst rates corresponding to the 90\% and 99\%
confidence upper limits on the normalization $R_H$ in each model are
given in Table~\ref{tbl:uphomo}.  The upper limits were determined by
finding the value of $R_H$ for which $\Delta \chi^2 = 2.7$ and $\Delta
\chi^2 = 6.6$, respectively, when $\chi^2$ is minimized with respect
to the other fit parameters (\cite{Avni76}).  In all cases, only a
modest fraction, 5--10\%, of the observed GRBs could come from a
homogeneous subpopulation (and thus from nearby SNe).  These upper
limits are comparable to the model-independent result found in the
previous paragraph.  They are slightly less constraining because of
the fact that our peak flux distribution refers only to the 1.024 s
time scale, so it takes no account of the paucity of faint bursts found
on the 4.096 and 8.192 s time scales (see section \ref{subsec:vvmax}).

These results were to be expected from the facts that 1) models D1,
D2, and D3 with power-law luminosity distributions already gave
excellent fits to the data without the presence of the homogeneous
($P^{-5/2}$) term, which is sharply peaked at low peak fluxes, and 2)
the fractional uncertainties on the rates in each bin are on the order
of 10\%.  The upper limits discussed here would be further reduced if
a given GRB must exhibit certain characteristics (e.g., single-peaked
time profile, lack of emission above 300 keV) in order to be
considered a candidate S-GRB (\cite{Bloom98}).  Norris, Bonnell, \&
Watanabe (1998) have found that only 0.25--0.5\% of BATSE GRBs have
temporal and spectral characteristics similar to GRB 980425.

\section{Discussion}
\label{sec:discuss}
The GRB peak flux distribution alone (on the 1.024 s time scale) only
weakly distinguishes between the non-evolving model $\rho(z) = $
constant (D1) and the evolving models where $\rho(z)$ is proportional
to an estimate of the star formation history (D2 and D3).  (In this
section we restrict our attention to models that include a power-law
distribution of intrinsic peak luminosities.)  A similar conclusion
has been reached previously by Krumholz et al.\ (1998), who analyze
the BATSE catalog data and find that to reliably distinguish the
non-evolving and evolving models requires data from more sensitive GRB
detectors and/or the measurement of more individual GRB redshifts.

The off-line sample of GRBs constitutes a more sensitive experiment
than the one analyzed by Krumholz et al.\ (1998).  Here we argue that
models similar (or identical) to D2 and D3, in which the GRB rate has
a significant peak in the redshift range $1 < z < 3$, {\it are
  modestly preferred over the constant rate density model}.  Two
independent lines of reasoning serve to denigrate the $\rho(z) =$
constant (``non-evolving'') models in favor of the evolving ones.

First, our search on the 1.024 s time scale can reach peak fluxes as
low as 0.16 ph cm$^{-2}$ s$^{-1}$ in the 50--300 keV band (50\%
detection efficiency).  But our searches on the 4.096 and 8.192 s time
scales are sensitive to peak fluxes (averaged over the matching time
scale) that are lower by factors of $\sim 2$ and $\sim 2\sqrt{2}$,
respectively, than the 1.024 s threshold.  In fact, most of the bursts
we detect have their highest signal-to-noise ratio in the 8.192 s
search.  Yet surprisingly few bursts are detected {\it exclusively\/}
on the longer 4.096 s and 8.192 s time scales.  This suggests that
there are relatively few faint GRBs waiting to be detected by a search
that is more sensitive than the one we carried out.  In this respect,
the evolving models (D2 and D3) appear to be more accurate.  They
predict that the number of bursts per logarithmic peak flux interval
will level off towards lower peak fluxes, and may even start to
decline (see Figure~\ref{fig:bestfit}).  The non-evolving model, on
the other hand, predicts that the number of bursts observed per
logarithmic peak flux interval will continue to increase towards lower
peak fluxes.

Since we have not derived peak fluxes on the 4.096 and 8.192 s time
scales in order to repeat the analysis of section
\ref{subsec:purecos}, we offer the following quantitative evidence
that the paucity of bursts detected on the longer time scales favors
the evolving models D2 and D3 over the non-evolving model D1.  On the
1.024 s time scale, the measured value of $\langle
(C_{min}/C_{max})^{3/2} \rangle = 0.247 \pm 0.006$ is trivially
consistent with the values of $\langle (P_{min}/P)^{3/2} \rangle$
found for the best-fit models in section \ref{subsec:purecos}.  But
the value of $\langle (C_{min}/C_{max})^{3/2} \rangle = 0.177 \pm
0.006$ found for {\it all\/} bursts detected by our search contains
information on the paucity of faint bursts on the 4.096 and 8.192 s
time scales.  We can compare it with the value of $\langle
(P_{min}/P)^{3/2} \rangle$ obtained by extrapolating the best-fit
models of section \ref{subsec:purecos} to the peak flux threshold
associated with the 8.192 s search.  Taking $0.18/(2 \sqrt{2}) = 0.06$
ph cm$^{-2}$ s$^{-1}$ as the approximate $P_{min}$ for the 8.192 s
search, we obtain the following values for $\langle (P_{min}/P)^{3/2}
\rangle$: 0.221 in model D1, 0.169 in model D2, and 0.147 in model D3.
Thus model D2 produces a value of $\langle (P_{min}/P)^{3/2} \rangle$
that is the most consistent with the value of $\langle
(C_{min}/C_{max})^{3/2} \rangle$ found for our full sample, and model
D1 produces the most inconsistent value.

Second, the inferred equivalent isotropic peak luminosities (in the
30--2000 keV range) of the 3 bursts for which associated redshifts
have been measured can be compared with the effective luminosity
distributions of the best-fit models.  The best-fit non-evolving model
(D1) predicts that 90\% of all GRBs should come from the narrow range
of intrinsic peak luminosities $(0.29$--$0.66)\ 10^{51} h^{-2}_{70}$
erg s$^{-1}$ (a factor of $\sim 2$).  The range from which 90\% of the
{\it observed\/} GRBs in this model are drawn is somewhat broader,
however: $(0.29$--$1.5)\ 10^{51} h^{-2}_{70}$ erg s$^{-1}$ (a factor
of $\sim$ 5).  In contrast, the intrinsic luminosities inferred from
the 3 bursts with associated redshift information span a much broader
peak luminosity range, $(0.6$--$37)\ 10^{51} h^{-2}_{70}$ erg s$^{-1}$
(a factor of $\sim 62$).  Bursts with peak luminosities as high as
those inferred for GRB 971214 ($(37 \pm 16)\ 10^{51} h^{-2}_{70}$ erg
s$^{-1}$) and GRB 980703 ($(2.2 \pm 0.4)\ 10^{51} h^{-2}_{70}$ erg
s$^{-1}$) are extremely rare events if the (non-evolving) model D1 is
correct.  On the other hand, the best-fit (evolving) models D2 and D3
allow a much broader (and uniformly higher) range of luminosities.
The best-fit model D2 predicts that 90\% of all GRBs are drawn from
the intrinsic peak luminosity range $(0.50$--$7.3)\ 10^{51}
h^{-2}_{70}$ erg s$^{-1}$ (a factor of $\sim 15$) and that 90\% of the
observed GRBs are drawn from the range $(0.53$--$72)\ 10^{51}
h^{-2}_{70}$ erg s$^{-1}$ (a factor of $\sim$ 130).  Likewise, the
best-fit model D3 predicts that 90\% of all GRBs are drawn from the
range $(1.5$--$9.3)\ 10^{51} h^{-2}_{70}$ erg s$^{-1}$ (a factor of
$\sim 6$) and 90\% of the observed GRBs are drawn from the range
$(1.5$--$40)\ 10^{51} h^{-2}_{70}$ erg s$^{-1}$ (a factor of $\sim$
25).  In the context of models D2 and D3, the GRBs 970508, 971214,
980613, and 980703 constitute a much more likely sample of detected
bursts than in model D1.

The effective redshift distributions (the rates of observed bursts as
a function of redshift) furnish another point of comparison with GRBs
970508, 971214, 980613, 980703.  As shown in Figure~\ref{fig:effzee} the
model D1 predicts a vanishingly small rate of observed bursts from the
redshift $z = 3.418$ measured for GRB 981214 (\cite{Kulkarni98a}).  If
model D1 were correct, then it would be remarkable that BATSE and {\it
  BeppoSAX\/} detected such a rare burst: only $1 \times 10^{-6}$ of
the rate distribution comes from higher redshifts.  On the other hand,
the evolving models D2 and D3 predict much higher rates of observed
bursts from $z = 3.418$.  Even in these models, however, such a high
redshift is exceptional: in model D2 only 0.4\% of the burst rate
distribution lies beyond $z = 3.4$, and in model D3 only 1.8\% does.
Still, as in the case of the inferred luminosity distributions, the 3
bursts with associated redshifts are a much more likely sample in
models D2 and D3, where the GRB rate follows an estimate of the star
formation history.

These results certainly do not prove that the GRB rate traces the star
formation rate.  The peak flux distributions alone fail to exclude the
non-evolving rate density model (D1) with high confidence, especially
in view of the unknown cosmological parameters which can be varied to
improve the fit.  Furthermore, until many more redshifts are
associated with specific GRBs and/or more sensitive GRB detectors go
on-line, the data will not be able to distinguish qualitatively
similar SFR evolution models such as D2 and D3.  Any evolution that
specifies a significant peak in the burst rate in the redshift range
$1 < z < 3$ is likely to be consistent with current data.  The
reasoning regarding the paucity of faint bursts detected {\it only\/}
on the 4 s and 8 s time scales should be addressed more quantitatively
in the context of the models; but this is difficult owing to our poor
understanding of the diverse time profiles of GRBs and of the
correlations between time profiles and peak fluxes.  Finally, we have
considered only 3 very specific ``straw-man'' models, and it may be
that none of them are particularly accurate representations of the
true GRB rate density and peak luminosity distributions.  For example,
a previous episode of star formation at high redshift could contribute
a hitherto undetected population of very faint GRBs.

Nevertheless, the results of our search appear to support the
conclusions reached by Totani (1997), Wijers et al.\ (1998), and
Krumholz et al.\ (1998), who showed that if the GRB rate traces the
SFR then relatively few faint ``classical'' GRBs are to be found below
the BATSE onboard detection threshold.  This information should be
useful to the designers and operators of future GRB detectors.

\begin{figure}
\plotone{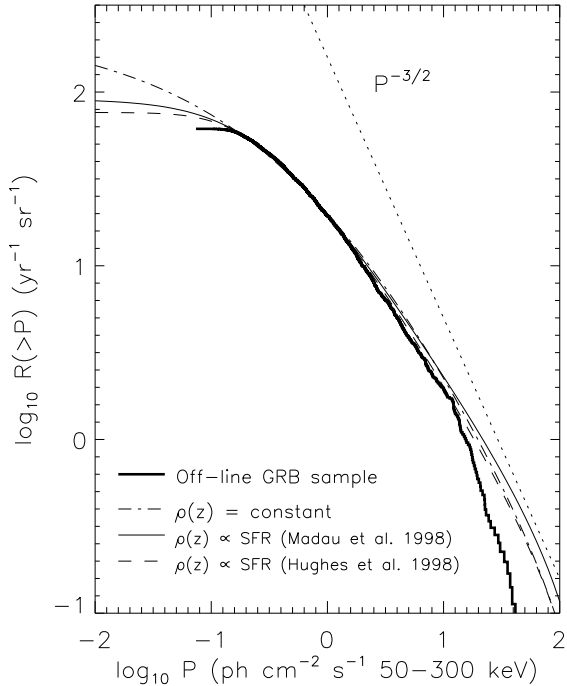}
\caption{Cumulative peak flux distributions for best-fit models.  The units
  of $R$ are bursts \protect{yr$^{-1}$ sr$^{-1}$}.  The observed peak
  flux distribution for the off-line sample is shown as the solid
  histogram.  The best-fit model D1 is shown as the dot-dashed line.
  Model D2 is shown as the solid line.  Model D3 is shown as the
  dashed line.  The discrepancy at high peak fluxes is not surprising
  because the fitting used only data from the interval 0.18--20.0 ph
  cm$^{-2}$ s$^{-1}$, but the curves in the figure are intended to
  reflect true cumulative distributions (with no lower or upper limit
  on peak flux).  The best-fit models appear to predict too many
  bright bursts, though the discrepancy between the data and the
  models is not statistically significant unless the comparison is
  restricted to bright bursts only (data above 10.0 ph cm$^{-2}$
  s$^{-1}$).  This suggests that the off-line sample should be
  combined with data from longer missions, such as {\it PVO\/} and/or
  {\it Ulysses} to check the behavior of the burst rate at high peak
  fluxes.}
\label{fig:cumrate}
\end{figure}

In Figure~\ref{fig:cumrate} we plot the cumulative rate distribution
of the off-line sample of GRBs along with the best fit models and
their extrapolations to lower and higher peak fluxes than are
available in the data.  The best fit models apparently predict too
many bright bursts above the maximum peak flux that was used to fit
the differential burst rate (20.0 ph cm$^{-2}$ s$^{-1}$).  This
discrepancy is not surprising because the fits did not include data at
high peak fluxes.  A Kolmogorov-Smirnov (K-S) test (\cite{Press95})
indicates that the discrepancy is not significant if the data and best
fit models are compared for {\it all\/} peak fluxes above 0.18 ph
cm$^{-2}$ s$^{-1}$.  (The probability of getting a larger value of the
K-S statistic is 0.4 for model D2.)  If the K-S test is restricted to
the data above 10.0 ph cm$^{-2}$ s$^{-1}$, however, the discrepancy is
significant (the probability of getting a larger value of the K-S
statictic is $6 \times 10^{-4}$ for model D2).  This means that the
extrapolations of the best fit models are inaccurate at high peak
fluxes.

Future research should explore the peak flux distribution over the
widest possible range.  For example, the GRB detector onboard {\it
  Pioneer Venus Orbiter\/} ({\it PVO}) operated for a much longer
mission than has BATSE (so far) and it has more completely sampled the
rate of very bright bursts.  When the peak fluxes of the GRBs detected
with {\it PVO\/} are calibrated to match the BATSE peak fluxes, it
would be of interest to see if the parameters of the best-fit models
found in this paper remain consistent with the number counts when the
very bright bursts are included.  Figure~\ref{fig:cumrate} shows that
the slopes of the best-fit models approach the $-3/2$ slope reported
to be consistent with the brightest {\it PVO\/} bursts
(\cite{Fenimore93}). 

The results of the best-fit models are otherwise generally consistent
with previous studies of the BATSE data.  In particular we find that
while we cannot constrain the full width of the power-law luminosity
function in any of the scenarios we considered, the best-fit models
yield intrinsic peak luminosity functions that contain 90\% of all
GRBs within a factor of 10--20.  This result (or a similar one) has
been previously obtained by Ulmer, Wijers, \& Fenimore (1995), Woods
\& Loeb (1995), Hakkila et al.\ (1996), and Horack et al.\ (1996).  We
find that the peak luminosity distribution of the observed bursts,
however, is wider: in model D2, 90\% of the observed bursts come from
a peak luminosity range that spans a factor of $\sim 100$ or more.  A
similar result is discussed by Loredo \& Wasserman (1998).

The effective redshift distributions that we obtain (see
Figure~\ref{fig:effzee}) are reasonably consistent with those obtained
by Krumholz et al.\ (1998) and Mao \& Mo (1998) using the BATSE
catalog data.  They are also consistent with limits on the redshifts
of GRB sources set by the non-detection of any gravitationally lensed
GRBs in the BATSE catalogs (\cite{Marani98}).  Holz, Miller, \&
Quashnock (1998) derive upper limits of $\langle z \rangle$ = 2.3
(68\% confidence) and $\langle z \rangle$ = 5.3 (95\% confidence) for
the average redshift of GRB sources in the Einstein-de Sitter
cosmology.  The values of $\langle z \rangle$ we obtain for our best
fit models (see Section \ref{subsec:purecos}) are all well within
these limits.  The effective redshift distributions are also
consistent with the disparity found by Norris et al.\ (1995) between
the duration distributions of bright and dim GRBs, which they
interpret as the signature of cosmological time dilation for a GRB
source distribution with the dimmest bursts at $z \approx 2$.

On the other hand, our effective redshift distributions are somewhat
at odds with the conclusions of Wijers et al.\ (1998) that the
faintest bursts observed with BATSE are at redshifts of $3 < z < 6$.
Part of the discrepancy is explained by the fact that rate model used
by Wijers et al.\ (1998) does not correct for the cosmological time
dilation of the co-moving burst rate.  According to our results with
models D1, D2, and D3, no more than 0.002\%, 1\%, or 5\%
(respectively) of the bursts observed with our off-line sample are
from redshifts larger than $z = 3.0$; and fewer than $8 \times
10^{-7}$\%, 0.07\%, and 0.3\%, respectively, are at redshifts greater
than $z = 4.0$.  So, it is possible that GRBs may be produced at
redshifts as high as $z \approx 6$; but even model D3, which allows
the highest redshifts, permits only a 7\% chance that one or more such
bursts have been observed among the 2265 in the off-line sample.

To summarize, our off-line search of archival BATSE data has explored
the distribution of GRB intensities at peak fluxes below the onboard
detection threshold.  We find a paucity of faint bursts detected on
the 4.096 and 8.192 s time scales that were not already detected on
the 1.024 s time scale.  The differential intensity distribution is
consistent with models in which the GRB rate traces the global star
formation history of the Universe; and it is marginally consistent
with the model in which the GRB rate is independent of redshift.  We
argue that the models in which the GRB rate traces the star formation
rate are nevertheless preferred, based on the paucity of faint bursts
detected exclusively on the 4.096 s and 8.192 s time scales, and on
the comparison of the inferred effective luminosity and redshift
distributions with the bursts for which redshifts have been measured.
As an application of the off-line GRB intensity distribution, we set a
limit of 10\% (99\% confidence) on the fractional rate of all GRBs
that could belong to a homogeneous (in Euclidean space) subpopulation
of burst sources (such as Type Ib/c supernovae).

\acknowledgments J. M. K. thanks Bob Rutledge for a careful reading of
the manuscript, and acknowledges useful discussions with Ed Fenimore,
Jon Hakkila, Gabriela Marani, Bob Nemiroff, and Ralph Wijers, as well
as support from NASA Graduate Student Researchers Program Fellowship
NGT8-52816.  W.  H. G.  L.  acknowledges support from NASA under grant
NAG5-3804. C. K.  acknowledges support from NASA under grants
NAG5-32490 and NAG5-4799.  J. v.\ P.  acknowledges support from NASA
under grants NAG5-2755 and NAG5-3674.

\clearpage


\clearpage

\begin{deluxetable}{lcc}
  \tablehead{\colhead{Detector/Mission} & \colhead{$\langle
      (C_{min}/C_{max})^{3/2} \rangle$} & \colhead{Reference}}
  \tablecaption{Values of $\langle (C_{min}/C_{max})^{3/2} \rangle$
    obtained by various GRB detectors.
\label{tbl:vvmax}}
\startdata
{\it PVO}                  &  $0.46 \pm 0.02$       &  Hartmann et
al. 1992 \nl
Konus/{\it Venera 11 \& 12} &  $0.45 \pm 0.03$       &  Higdon \&
Schmidt 1990 \nl
A4/{\it HEAO 1}            &  $0.40 \pm 0.08$       &  Schmidt,
Higdon, \& Hueter 1988 \nl
GRS/{\it SMM}              &  $0.40 \pm 0.025$      &  Matz et al.\ 1992 \nl
GBD/{\it Ginga}            &  $0.35 \pm 0.035$      &  Ogasaka et
al. 1991 \nl
BATSE/{\it CGRO} (3B)      &  $0.33 \pm 0.01$       &  Meegan et
al. 1996 \nl
BATSE/{\it CGRO} (off-line, & ~ & ~ \nl
~~1.024 s search only) & $0.247 \pm 0.006$ & (this paper) \nl
BATSE/{\it CGRO} (all off-line) & $0.177 \pm 0.006$ & (this paper) \nl
\enddata
\end{deluxetable}

\scriptsize
\begin{deluxetable}{cccc}
  \tablehead{\colhead{$P_1$} & \colhead{$P_2$} &
  \colhead{$\Delta N_{obs}$} & \colhead{$\Delta R$ (yr$^{-1}$ sr$^{-1}$)}}
  \tablecaption{Data for fitting differential peak flux distribution
\label{tbl:pfdata}}
\startdata
 0.180 & 0.202 &  87  &  2.45 $\pm$ 0.26 \nl
 0.202 & 0.227 &  83  &  2.34 $\pm$ 0.26 \nl
 0.227 & 0.254 &  99  &  2.80 $\pm$ 0.28 \nl
 0.254 & 0.285 &  111 &  3.13 $\pm$ 0.30 \nl
 0.285 & 0.320 &  92  &  2.60 $\pm$ 0.27 \nl
 0.320 & 0.359 &  96  &  2.71 $\pm$ 0.28 \nl
 0.359 & 0.403 &  95  &  2.68 $\pm$ 0.27 \nl
 0.403 & 0.452 &  114 &  3.22 $\pm$ 0.30 \nl
 0.452 & 0.507 &  93  &  2.62 $\pm$ 0.27 \nl
 0.507 & 0.569 &  79  &  2.23 $\pm$ 0.25 \nl
 0.569 & 0.639 &  82  &  2.31 $\pm$ 0.26 \nl
 0.639 & 0.717 &  91  &  2.57 $\pm$ 0.27 \nl
 0.717 & 0.804 &  73  &  2.06 $\pm$ 0.24 \nl
 0.804 & 0.902 &  83  &  2.34 $\pm$ 0.26 \nl
 0.902 & 1.000 &  52  &  1.47 $\pm$ 0.20 \nl
 1.000 & 1.259 &  117 &  3.30 $\pm$ 0.31 \nl
 1.259 & 1.584 &  111 &  3.13 $\pm$ 0.30 \nl
 1.584 & 1.995 &  104 &  2.93 $\pm$ 0.29 \nl
 1.995 & 2.511 &  72  &  2.03 $\pm$ 0.24 \nl
 2.511 & 3.162 &  59  &  1.66 $\pm$ 0.22 \nl
 3.162 & 3.981 &  48  &  1.35 $\pm$ 0.20 \nl
 3.981 & 5.011 &  40  &  1.13 $\pm$ 0.18 \nl
 5.011 & 6.309 &  27  &  0.76 $\pm$ 0.15 \nl
 6.309 & 7.943 &  26  &  0.73 $\pm$ 0.14 \nl
 7.943 & 20.00 &  64  &  1.80 $\pm$ 0.23 \nl
\enddata
\end{deluxetable}
\normalsize

\begin{deluxetable}{cccc}
  \tablehead{\colhead{GRB} & \colhead{\begin{tabular}{c}Peak
        Flux\\(50--300 keV)\end{tabular}} & \colhead{$z$} &
    \colhead{\begin{tabular}{c}Implied $L$\\($10^{51} h^{-2}_{70}$ erg
        s$^{-1}$ in 30--2000 keV)\end{tabular}}}
  \tablecaption{Redshifts and implied equivalent isotropic peak
    luminosities \label{tbl:rlums}}
\startdata
970508    &   $0.97 \pm 0.05$\tablenotemark{a} &
        0.835\tablenotemark{c} & $0.6 \pm 0.1$ \nl
971214    &   $1.95 \pm 0.05$\tablenotemark{a} &
        3.418\tablenotemark{d} & $37 \pm 16$ \nl
980613    &   $0.63 \pm 0.05$\tablenotemark{b} &
        1.096\tablenotemark{e} & $0.8 \pm 0.2$ \nl
980703    &   $2.39 \pm 0.06$\tablenotemark{a} &
        0.966\tablenotemark{f} & $2.2 \pm 0.4$ \nl
\enddata
\tablenotetext{a}{Meegan et al.\ (1998)}
\tablenotetext{b}{Woods, Kippen, \& Connaughton (1999)}
\tablenotetext{c}{Metzger et al.\ (1997)}
\tablenotetext{d}{Kulkarni et al.\ (1998a)}
\tablenotetext{e}{Djorgovski et al.\ (1999)}
\tablenotetext{f}{Djorgovski et al.\ (1998)}
\end{deluxetable}

\begin{deluxetable}{cccc}
  \tablehead{\colhead{$\rho(z)$} & 
    \colhead{\begin{tabular}{cc}$R_0$\\($h^{3}_{70}$ Gpc$^{-3}$ yr$^{-1}$ sr$^{-1}$)\end{tabular}} &
    \colhead{\begin{tabular}{cc}$L_{0}$\\($10^{51}\ h^{-2}_{70}$ erg s$^{-1}$)\end{tabular}} &
    \colhead{\begin{tabular}{cc}$\chi^2$\\(23 d.o.f)\end{tabular}}}
  \tablecaption{Best fit parameters for monoluminous cosmological models
\label{tbl:fitresstd}}
\startdata
D1 (constant) & $9.45^{+0.39}_{-0.78}$ & $0.40^{+0.06}_{-0.02}$ & 32.3 \nl
D2 (SFR)      & $2.0 \pm 0.1$ & $1.5 \pm 0.1$ & 64.1 \nl
D3 (AGN)      & $1.9 \pm 0.1$ & $3.1^{+0.4}_{-0.2}$ & 27.8 \nl
\enddata
\end{deluxetable}

\begin{deluxetable}{ccccc}
  \tablehead{\colhead{$\rho(z)$} & 
    \colhead{\begin{tabular}{cc}$R_0$\\($h^{3}_{70}$ Gpc$^{-3}$ yr$^{-1}$ sr$^{-1}$)\end{tabular}} &
    \colhead{\begin{tabular}{cc}$L_{min}$\\($10^{51}\ h^{-2}_{70}$ erg s$^{-1}$)\end{tabular}} &
    \colhead{$\beta$} &
    \colhead{\begin{tabular}{cc}$\chi^2$\\(22 d.o.f)\end{tabular}}}
  \tablecaption{Best fit parameters for cosmological models with
    a power-law luminosity function
\label{tbl:fitrespl}}
\startdata
D1 (constant) & $8.8 \pm 1.3$ & 0.29$^{+0.08}_{-0.06}$ & $4.6_{-1.4}$\tablenotemark{a} & 32.9 \nl
D2 (SFR)      & $2.2 \pm 0.3$ & 0.48$^{+0.20}_{-0.10}$ & $2.1^{+0.3}_{-0.2}$ & 17.9 \nl
D3 (AGN)      & $2.1 \pm 0.2$  & 1.44$^{+0.53}_{-0.40}$ & $2.6^{+1.0}_{-0.4}$ & 17.4 \nl
\enddata
\tablenotetext{a}{This parameter is not constrained above the the best
  fit value when the other parameters are free to vary.}
\end{deluxetable}

\begin{deluxetable}{ccc}
  \tablehead{\colhead{$\rho(z)$} & 
    \colhead{\begin{tabular}{c}$\rho_0$\\($h^{3}_{70}$ Gpc$^{-3}$ yr$^{-1}$)\end{tabular}} &
    \colhead{\begin{tabular}{c}GEM\tablenotemark{a}\\(Myr$^{-1}$)\end{tabular}}}
  \tablecaption{Co-moving $z=0$ burst rate for monoluminous cosmological models
\label{tbl:ratestd}}
\startdata
D1 (constant) & $11.9^{+0.5}_{-1.0}$ & $\sim$ 2 \nl
D2 (SFR)      & $0.78 \pm 0.04$ & $\sim$ 0.2\nl
D3 (AGN)      & $0.46 \pm 0.02$ & $\sim$ 0.1\nl
\enddata
\tablenotetext{a}{Galactic Events per Million years.}
\end{deluxetable}

\begin{deluxetable}{ccc}
  \tablehead{\colhead{$\rho(z)$} & 
    \colhead{\begin{tabular}{cc}$\rho_0$\\($h^{3}_{70}$ Gpc$^{-3}$ yr$^{-1}$)\end{tabular}} &
    \colhead{\begin{tabular}{c}GEM\tablenotemark{a}\\(Myr$^{-1}$)\end{tabular}}}
  \tablecaption{Co-moving $z=0$ burst rate for cosmological models with
    a power-law luminosity function
\label{tbl:ratepl}}
\startdata
D1 (constant) & $11.1 \pm 1.6$ & $\sim$ 2 \nl
D2 (SFR)      & $0.87 \pm 0.12$ & $\sim$ 0.2 \nl
D3 (AGN)      & $0.51 \pm 0.05$ & $\sim 0.1$ \nl
\enddata
\tablenotetext{a}{Galactic Events per Million years.}
\end{deluxetable}

\normalsize
\begin{deluxetable}{ccc}
  \tablehead{\colhead{$\rho(z)$} & 
    \colhead{90\%} &
    \colhead{99\%}}
  \tablecaption{Upper limits on fractional GRB rate due to a possible
    homogeneous (in Euclidean space) sub-population of GRBs
\label{tbl:uphomo}}
\startdata
D1 (constant) & 5.1\% & 6.2\% \nl
D2 (SFR)      & 6.9\% & 10.0\% \nl
D3 (AGN)      & 7.7\% & 11.2\% \nl
\enddata
\end{deluxetable}

\end{document}